\documentclass[11pt,twoside]{article}  
  
\usepackage{asp2006}  
\usepackage{epsf}  
\usepackage{lscape}  
\usepackage{graphicx}
\begin{document}  
\title{Non-linear pulsations in Wolf-Rayet stars}   
\author{Sebastian Wende, Wolfgang Glatzel, Sonja Schuh}   
\affil{Institut f\"ur Astrophysik, Georg-August-Universit\"at G\"ottingen,
  Friedrich-Hund-Platz~1, D-37077 G\"ottingen, Germany}    
  
\begin{abstract} 
Numerical simulations of the evolution of strange-mode instabilities into the
			     non-linear regime have been performed for a wide range of stellar parameters
			     for Wolf-Rayet stars. It has been shown that the Wolf-Rayet models reach radial
			     velocities which amount to up to $30\%$ of their escape velocity. The acoustic
			     luminosities suggest a connection to the observed mass loss. Most of the models show a
			     \emph{jump} in the mean effective temperature after reaching the non-linear
			     regime. This \emph{jump} is related to the run of the opacity.
\end{abstract}  
  
  
To determine the fate of
		unstable stars, the evolution of their instabilities has to be
		followed into the non-linear regime by numerical simulations
		\citep{Grott2005}. Linear
		stability analyses have shown that strong
		instabilities with dynamical growth rates prevail in many types of
		massive stars. These are associated with so
		called \emph{strange modes} \citep{GG1990}. Models of Wolf-Rayet stars exhibit the
		strongest \emph{strange mode} instabilities found so far \citep{Glatzel1993}. Thus they are promising
		candidates for a non-linear stability analysis.

The Wolf-Rayet
		models investigated reach final radial velocity amplitudes in the
		non-linear regime of the order of $10^7cm/s$. These values are
		comparable with the escape velocity. To investigate a possible
		 		connection between stellar instabilities and a
		 		pulsationally driven mass loss, the acoustic
		 		energy flux is considered. It represents the
		 		mechanical energy transferred by shock waves to the star's
		 		atmosphere \citep{Grott2005}. We emphasize
		                that the acoustic luminosity and thus any mass
		                loss rate based on it strongly depends on the
		                artificial viscosity which has to be used in the
		                simulations. However, since viscosity has a dissipative
		                effect the derived mass loss rates have to be
		                regarded as lower limits (see Table 1).
	      
	      \begin{table}[!ht]			
	       \begin{center}
		\begin{scriptsize}
		\caption{{{Acoustic luminosities and mass loss
		rates based on them for models with different initial
		effective temperatures. For the models, the mass-luminosity
		relation of the He-ZAMS was adopted.}}} 
		 \begin{tabular}{cccccccc}
		  \tableline
		  \noalign{\smallskip}
		  $M$           & $T_{initial}$ & $L_{acou.}$  & $\dot{M}$
		   &\hspace{1cm}
		   $M$           & $T_{initial}$ & $L_{acou.}$  & $\dot{M}$ \\ 
		  $[M_{\odot}]$ & $[K]$         & $[erg/s]$    &
		  $[M_{\odot}/a]$ & \hspace{1cm}$[M_{\odot}]$ & $[K]$         & $[erg/s]$    & $[M_{\odot}/a]$\\
		  \noalign{\smallskip}
		  \tableline
                  \noalign{\smallskip}              
		  $7.021$ &$ 60000$ &$ 5.00\cdot10^{33}$ &$ 1.40\cdot10^{-8} $&\hspace{1cm}$ 13.040 $&$ 60000 $&$ 8.17\cdot10^{34} $&$ 2.30\cdot10^{-7}$\\ 
		 $ 7.021$ & $90000$ & $2.55\cdot10^{32}$ &$ 3.17\cdot10^{-10}$&\hspace{1cm}$ 13.040 $&$ 90000 $&$ 6.78\cdot10^{33} $&$ 8.44\cdot10^{-9} $ \\ 
		 $ 9.027 $&$ 60000 $&$ 2.64\cdot10^{34} $&$ 7.50\cdot10^{-8} $&\hspace{1cm}$ 17.050 $&$ 60000 $&$ 6.84\cdot10^{34} $&$ 1.87\cdot10^{-7} $\\ 
		 $ 9.027 $&$ 90000 $&$ 1.54\cdot10^{33} $&$ 1.95\cdot10^{-9} $&\hspace{1cm}$ 17.050 $&$ 90000 $&$ 1.95\cdot10^{35} $&$ 2.37\cdot10^{-7}$\\ 
		 
		  \noalign{\smallskip}
		  \tableline
		\end{tabular}
	       \end{scriptsize}
	       \end{center}
	       \end{table}
	    
		
		After reaching the non-linear regime all unstable
		models exhibit a \emph{jump} in the mean effective
		temperature. 
                The final mean effective temperature
		is independent of mass and largely independent of the initial
		effective temperature. We find only two final mean effective
		temperatures around $30000K$ and $15000K$ respectively. 
		Models with an initial effective temperature above $\approx
		32000K$ reach a mean effective temperature around $30000K$,
		models with initial temperatures below this value end up with a mean
		effective temperature around $15000K$.
		Considering the run of the opacity,
		  these two temperatures correspond to the descending branches 
		  of the opacity from the respective maximum towards lower
		temperatures (see Fig.1). 
                Should descending branches of the opacity in general give rise
		to a range of possible final mean temperatures of pulsating
		models, a third range above $\approx 100000K$
		  would be expected. In fact, some
		  models show a \emph{jump} to this temperature. The final
		mean position of the pulsating models in the
		  HRD is of particular interest. Most of the models investigated
		  move to the right and settle in a narrow temperature
		  strip (Fig.1). It is remarkably close to the observed positions of
		  Wolf-Rayet stars.

 \begin{figure}[!ht]
 \begin{center}
  \includegraphics[width=0.42\textwidth,angle=270]{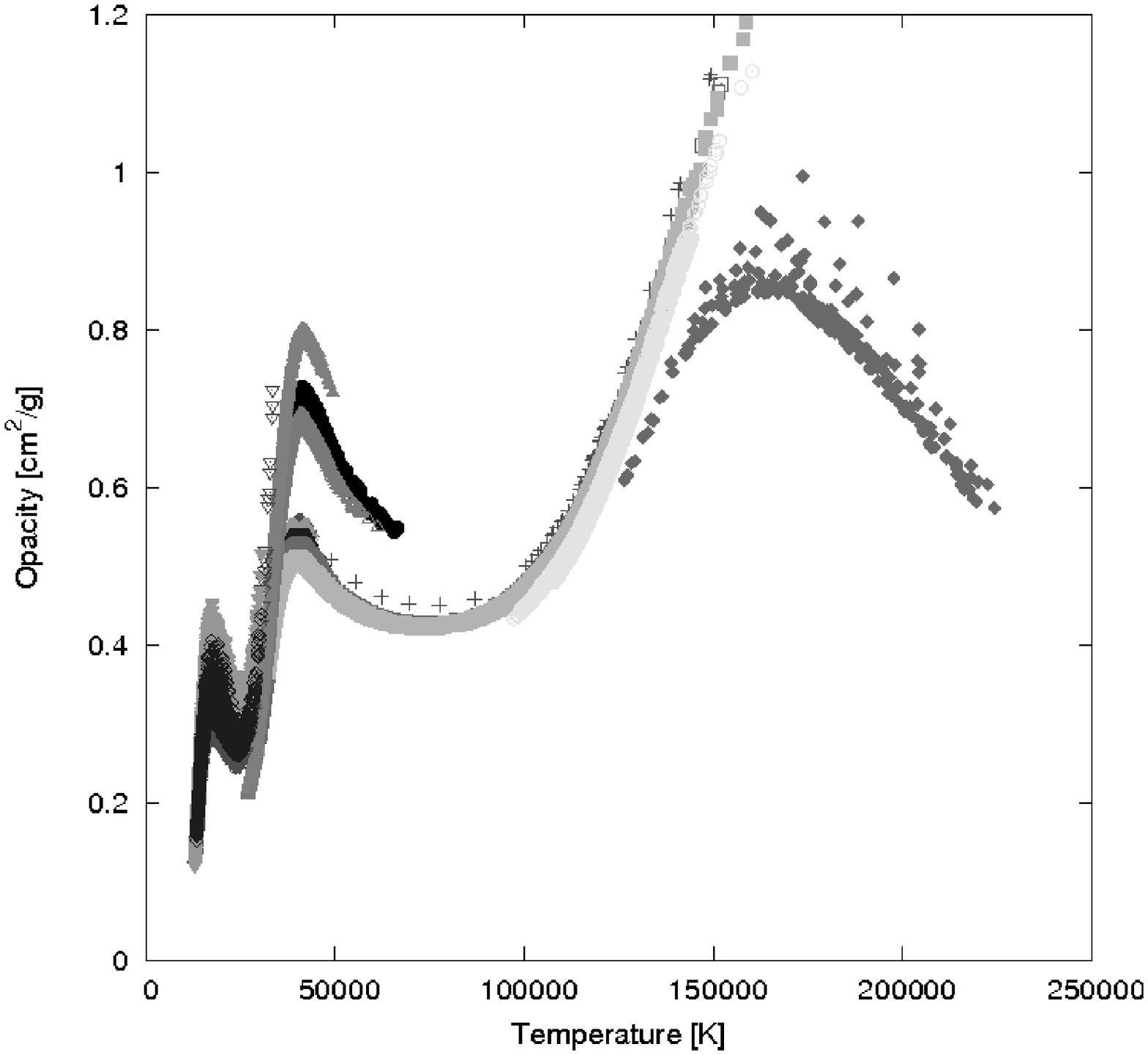}
  \includegraphics[width=0.42\textwidth,angle=270]{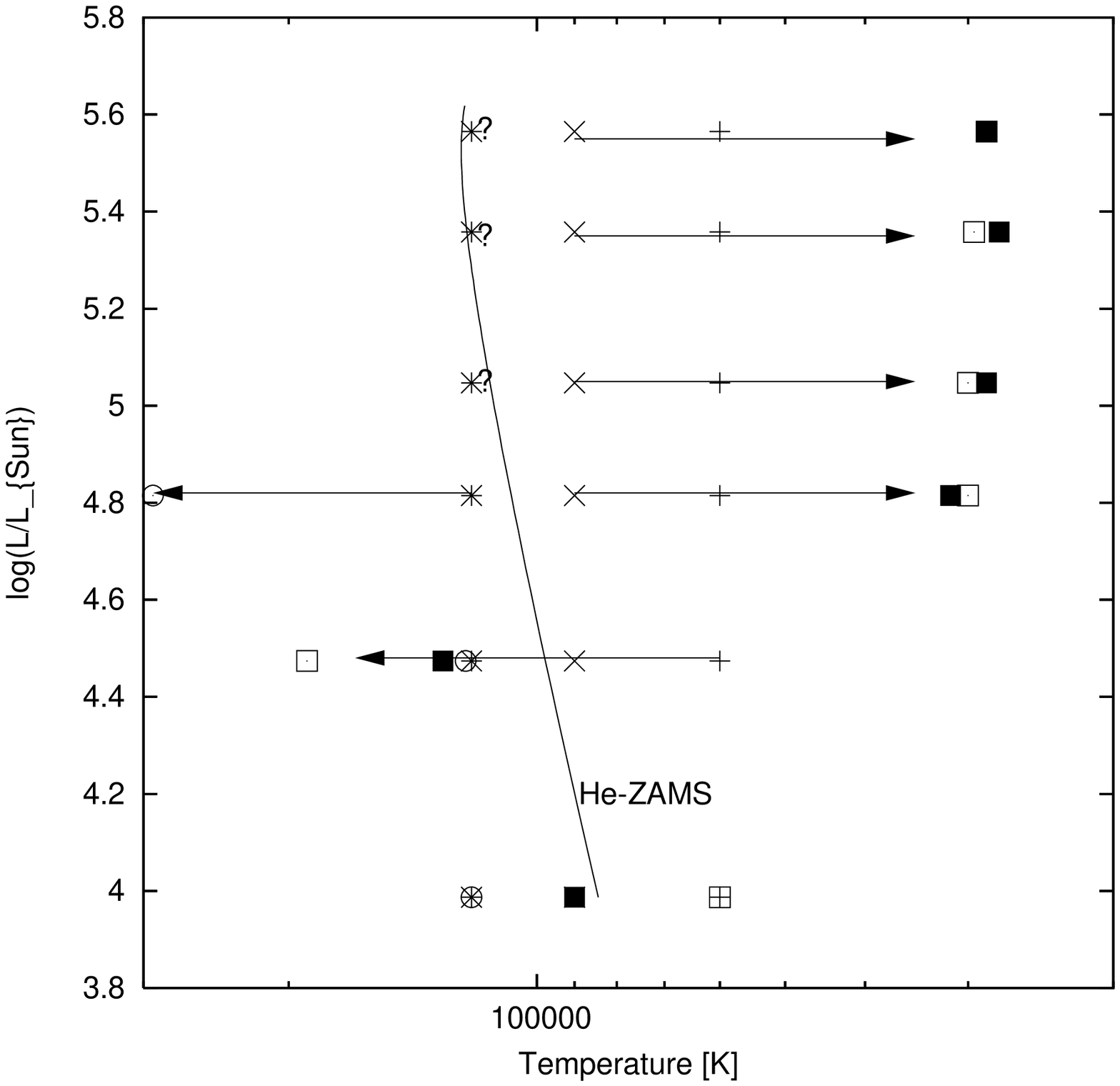}
\caption{{{Left: Run of the opacity in the stellar envelope as a
		function of temperature. Different stellar models were used
		(different gray scales correspond to different models) to cover a wide range of
		temperatures. For further details see text. Right: Hertzsprung-Russell-Diagram. 
                Indicated are the positions of the initial models by crosses
		and 
                asterisks and the final mean positions by squares and circles. The mean luminosities 
		stay constant. Arrows indicate the direction of the
		evolution. For models with question marks, we were not able
		to determine a final mean temperature. Models without arrows don't change
		their mean temperature. The He-ZAMS is shown for comparison.}}}\end{center}
\end{figure}

\acknowledgements 
Financial support by the Institut f\"ur Astronomie und Astrophysik
T\"ubingen is gratefully acknowledged.   
  

\end{document}